\documentclass[superscriptaddress, prd, aps,amsmath,amssymb,showpacs,showkeys, twocolumn]{revtex4-2}
\usepackage[dvips]{graphicx,color}
\usepackage{subfig}
\usepackage{times}
\usepackage{xcolor}
\usepackage[%
  colorlinks=true,
  urlcolor=blue,
  linkcolor=red,
  citecolor=blue
]{hyperref}
\usepackage{orcidlink}

\begin{document}
\title{Constraints on Quasinormal modes from Black Hole Shadows in regular non-minimal Einstein Yang-Mills Gravity}

\author{Dhruba Jyoti Gogoi\orcidlink{0000-0002-4776-8506}}
\email[Email: ]{moloydhruba@yahoo.in}

\affiliation{Department of Physics, Moran College, Moranhat, Charaideo 785670, Assam, India}
\affiliation{Theoretical Physics Division, Centre for Atmospheric Studies, Dibrugarh University, Dibrugarh
786004, Assam, India}

\author{Supakchai Ponglertsakul}
\email[Email: ]{supakchai.p@gmail.com}
\affiliation{Strong Gravity Group, Department of Physics, Faculty of Science, Silpakorn University, Nakhon Pathom 73000, Thailand}

%\date{}
\begin{abstract}
This work deals with the scalar quasinormal modes using higher order WKB method and black hole shadow in non-minimal Einstein Yang-Mills theory. To validate the results of quasinormal modes, time domain profiles are also investigated. We found that with an increase in the magnetic charge of the black hole, the ring-down gravitational wave increases non-linearly and damping rate decreases non-linearly. The presence of magnetic charge also results in a decrease in the black hole shadow non-linearly. It is found that for large values of the coupling parameter, the black hole changes to a solitonic solution and the corresponding ring-down gravitational wave frequency increases slowly with a decrease in the damping rate. For the solitonic solutions, the shadow is also smaller. The constraints on the model parameters calculated using shadow observations of M87* and Sgr A* and an approximate analytic relation between quasinormal modes and shadow at the eikonal limit is discussed.
\end{abstract}

%\pacs{04.30.Tv, 04.50.Kd}

\maketitle
\section{Introduction}\label{sec01}

Black holes are one of the most fundamental objects in the universe. They originate from collapsing stars and are governed by Einstein's theory of general relativity (GR). Regardless of complex interior structure of dying stars, once formed, black holes become one of the simplest celestial objects. According to the no-hair theorem, black holes are uniquely described by three physical parameters i.e, mass, angular momentum and electric charge. However, there are several scenarios where the no-hair theorem can be evaded. For instance, in the Einstein-Yang-Mills (EYM) theory and its variants, several numerical black holes with Yang-Mills hair are studied \cite{PhysRevLett.64.2844, *Aichelburg:1992st, *Donets:1992zb, *Kleihaus:1997rb, *Winstanley:1998sn, *PhysRevD.93.064064, *PhysRevD.47.2242, *Ponglertsakul:2016fxj}. When asymptotic structure of spacetime is modified to boxlike boundary, various hairy black holes are found \cite{Dolan:2015dha,*Ponglertsakul:2016wae,*Ponglertsakul:2016anb, *PhysRevLett.116.141101, *PhysRevLett.116.141102, *Sanchis-Gual:2016tcm, *Basu:2016srp, *Peng:2017squ, *Dias:2018yey}. 

Moreover, there has been much interest in gravitational theories coupled with nonlinear electrodynamics sources. Such marriage leads to a novel type of black hole with event horizon but possesses no essential singularity or regular black hole \cite{Ayon-Beato:1998hmi,*PhysRevD.70.047504,*PhysRevD.90.124045,*Ma:2015gpa,*Bronnikov:2000vy,*Fan:2016hvf}. Regular black holes are also studied in modified theories of gravity, for instance, $f(R)$ gravity \cite{Rodrigues:2015ayd,Nojiri:2017kex}, Gauss-Bonnet gravity \cite{Nojiri:2017kex,PhysRevD.97.104050,Estrada:2023cyx}. An exact spherically symmetric Wu-Yang monopole solution is derived and studied \cite{Balakin:2006gv,WuYang,Shnir}. This is a non-minimally extended of Einstein-Yang-Mills equation with $SU(2)$ Wu-Yang ansatz. Later, this regular magnetic black hole is generalized to include a cosmological constant \cite{Balakin:2015gpq}.

As predicted by GR, any accelerated massive objects create disturbances in spacetime around them. The disturbances propagate through spacetime in the form of gravitational waves (GWs). These waves contain crucial information about their astrophysical sources. When two black holes collide and merge into a single massive black hole, during this process GW is emitted continuously. After the merger, a single black hole emits GWs signal which is prominently dominated by a ringdown mode. The black hole ringdown is generally described by quasinormal modes (QNMs) \cite{Vishveshwara:1970zz, Press:1971wr,Kokkotas:1999bd,Konoplya:2011qq}. An associated complex frequency is determined by black hole's mass and angular momentum. The real component of the frequency is the emission frequency while the imaginary component denotes exponential decay. Thus, studying QNMs of black holes is crucial since they allow us to attain various characteristics of black holes.

Remarkably, the detection of GWs \cite{PhysRevLett.116.061102} in 2015 has opened a new window to testing theories of gravity. Vast number of works have been devoted to study the QNMs of various black holes. QNMs of nonminimally coupled scalar field in three dimensional spacetime are explored in \cite{Rincon:2018ktz}. In addition, QNMs of nonminimally coupled scalar field in a five dimensional Einstein-Power-Maxwell background are studied via the Wentzel–Kramers–Brillouin (WKB) method and pseudospectral Chebyshev method \cite{Rincon:2021gwd}. Optical properties and QNMs of black holes dependence on the Yang-Mill charges are investigated in Einstein-Power-Yang-Mills theory \cite{Gogoi:2023ffh}. QNMs of black holes within the framework of generalized uncertainty principle (GUP) are analyzed \cite{Karmakar:2022idu,Lambiase:2023hng,Anacleto:2021qoe,Gogoi:2022wyv}. Due to variety of black hole solutions in modified theories of gravity, there are many works considering QNMs of black holes in various models e.g., $f(Q)$ gravity \cite{Gogoi:2023kjt}, $f(R,T)$ gravity \cite{Tangphati:2023xnw}, Horndeski gravity \cite{Tattersall:2018nve,Sekhmani:2023ict}, Rastall gravity \cite{Gogoi:2021dkr,Gogoi:2021cbp,Gogoi:2023lvw} and de Rham-Gabadadze-Tolley (dRGT) massive gravity \cite{Burikham:2017gdm,Ponglertsakul:2018smo,Wuthicharn:2019olp,Wongjun:2019ydo,Hendi:2021yxi}. Very recently, QNMs of black hole in emergent gravity framework are computed via the WKB and asymptotic iteration methods \cite{Gogoi:2023fow}.

In recent times, the visual representation of black holes has garnered significant attention, driven by observations of celestial objects like M87$^\star$ \cite{EventHorizonTelescope:2019dse,EventHorizonTelescope:2019uob,EventHorizonTelescope:2019jan,EventHorizonTelescope:2019ths,EventHorizonTelescope:2019pgp,EventHorizonTelescope:2019ggy} and Sgr A$^\star$ \cite{EventHorizonTelescope:2022wkp}. These observations have sparked interest in the realms of both GR and modified gravity (MOG), offering an intriguing opportunity to depict and analyze hypothetical images associated with these observed phenomena. This avenue allows us to engage in a geometric and topological exploration, comparing the theoretical depiction of black hole images with the actual images of M87$^\star$ and Sgr A$^\star$. Consequently, the visualization of black holes assumes a captivating optical dimension, particularly as we endeavour to uncover the relationships between optical features, such as the shadows they cast, through the application of thermodynamics \cite{Cai:2021uov,Zhang:2019glo} and quasi-normal modes \cite{Jusufi:2020dhz,Yang:2021zqy}. Additionally, when considering the shape data, the shadow images can take the form of either distorted geometries on the image plane \cite{Grenzebach:2014fha,Abdujabbarov:2016hnw,Amarilla:2011fx,Kumar:2020owy,Belhaj:2022kek} or a pattern of concentric circles, contingent on the presence of a rotation parameter within the spacetime. To gain a deeper understanding of these shadow behaviours, the Hamilton-Jacobi formalism \cite{Carter:1968rr} emerges as a compatible analytical framework. It is based on the concept that massless photons in the vicinity of a black hole generate specific orbits around the region, a phenomenon referred to as the geodesic null background. In the realm of black hole imaging, ongoing research aims to provide a comparative analysis of observed images, such as those of M87$^\star$ and Sgr A$^\star$, in terms of their size and shape.

%and observation of black hole shadow \cite{EventHorizonTelescope:2019dse} in 2019 has opened a new window to testing theories of gravity. 

Motivated by these studies, in this work, we aim to investigate the QNMs and black hole shadows in the framework of regular non-minimally EYM theory in four-dimensional spacetime. We also introduce constraints on the model parameters from black hole shadow observations and try to relate the shadow behaviour with the QNMs of the black hole spacetime.

This paper is organized as follows. 
In Section \ref{sec02}, we discuss the EYM theory and the associated black hole solution in brief. In Section \ref{sec03}, massless scalar perturbations on the black hole spacetime have been investigated and the QNMs spectrum is analysed. Section \ref{sec04} deals with the evolution of scalar perturbation. In Section \ref{sec05}, we investigate the black hole shadow. In Section \ref{sec06}, we discuss how QNMs are linked with shadows of the black hole. Finally in Section \ref{sec07}, we provide a concluding remark of our work.

Throughout the manuscript, we consider $c=\hbar = 8\pi G = 1$.

\section{Regular non-minimal Einstein Yang-Mills Black Hole}\label{sec02}
In this section, we shall discuss the theory and the black hole solution in brief. The associated action of the $4$-D regular non-minimally EYM theory is \cite{Balakin:2006gv, Balakin:2015gpq}%
\begin{equation}
S=\frac{1}{8\pi }\int d^{4}x\sqrt{-g}\left[ R+\frac{1}{2}F_{\mu \nu
}^{\left( a\right) }F^{\mu \nu {\left( a\right) }}+\frac{1}{2}\bar{R}^{\alpha
\beta \mu \nu }F_{\alpha \beta }^{\left( a\right) }F_{\mu \nu }^{\left(
a\right) }\right] ,
\end{equation}%
where $R$ stands for the Ricci curvature scalar. The indices labeled with Greek letters range from 0 to 3, while those labeled with Latin letters range from 1 to 3. Additionally, the notation $F_{\mu \nu }^{\left( a\right) }$ represents the Yang-Mills (YM) tensor and is connected to the YM potential $A_{\mu }^{\left( a\right) }$ through the equation:
\begin{equation}
F_{\mu \nu }^{\left( a\right) }=\nabla _{\mu }A_{\nu }^{\left( a\right)
}-\nabla _{\nu }A_{\mu }^{\left( a\right) }+f_{\left( b\right) \left(
c\right) }^{\left( a\right) }A_{\mu }^{\left( b\right) }A_{\nu }^{\left(
c\right) },
\end{equation}%
where $\nabla {\mu }$ represents the covariant derivative and the symbols $f{\left( b\right) \left( c\right) }^{\left( a\right) }$ denote the real structure constants of the three-parameter YM gauge group $SU\left( 2\right) $. The tensor $\bar{R}^{\alpha \beta \mu \nu }$ is defined as \cite{Balakin:2015gpq}
\begin{align}
\bar{R}^{\alpha \beta \mu \nu } &=\frac{\xi _{1}}{2}R\left( g^{\alpha \mu }g^{\beta \nu
}-g^{\alpha \nu }g^{\beta \mu }\right) \nonumber \\
&~~~~~+ \frac{\xi _{2}}{2}\left( R^{\alpha \mu }g^{\beta \nu }-R^{\alpha \nu
}g^{\beta \mu }+R^{\beta \nu }g^{\alpha \mu }-R^{\beta \mu }g^{\alpha \nu
}\right) \nonumber \\
&~~~~~+\xi _{3}R^{\alpha \beta \mu \nu }.
\end{align}

Here, $R^{\alpha \beta }$ and $R^{\alpha \beta \mu \nu }$ represent the Ricci and Riemann tensors, respectively, and $\xi _{i}$ $(i=1,2,3)$ denote the non-minimal coupling parameters between the YM field and the gravitational field. Assuming the gauge field is described by the Wu-Yang ansatz and taking $\xi _{1}=\xi ,\xi _{2}=4\xi ,\xi _{3}=-6\xi $ with $\xi >0$, a regular, static, and spherically symmetric black hole solution was discovered \cite{Balakin:2006gv, Balakin:2015gpq}. This black hole solution is described by the metric \cite{Balakin:2006gv, Balakin:2015gpq}, 
\begin{equation}
ds^{2}=-f\left( r\right) dt^{2}+f^{-1}(r)dr^{2}+r^{2}\left( d\theta ^{2}+\sin
^{2}\theta d\phi ^{2}\right), \label{M1}
\end{equation}%
where the metric function has the following form:
\begin{equation}
f\left( r\right) =1+\left( \frac{r^{4}}{r^{4}+2\xi Q^{2}}\right) \left( 
\frac{Q^{2}}{r^{2}}-\frac{2M}{r}\right) ,
\end{equation}%
with $M$ representing the mass of the black hole and $Q$ being the magnetic charge.

The electromagnetic field four-potential for the regular non-minimal magnetic black hole is given by:
\begin{equation}
A_{\nu }=\left( 0,0,0,Q\cos \theta \right) .
\end{equation}
It's evident that when $Q=0$, the metric (\ref{M1}) simplifies to the standard Schwarzschild black hole metric, while when $\xi =0$, it corresponds to the Reissner-Nordstr\"om black hole metric but with a magnetic charge instead of an electric charge. The quantity of event horizons of the non-minimal EYM black hole depends on both the non-minimal parameter and the magnetic charge, potentially resulting in multiple horizons. In this work, we shall use the black hole line element \eqref{M1} to study the massless scalar QNMs and the black hole shadow.

\section{Massless Scalar Quasinormal modes}
\label{sec03}

In this section, we shall investigate the QNMs associated with massless scalar perturbation. In this case, we shall assume that the test field has negligible impact or influence on the black hole spacetime defined in the previous section and the associated scalar field is massless in nature. We obtain a Klein-Gordon-type equation associated with the QNMs while considering the pertinent conservation relations associated with the spacetime under consideration. We shall implement Pad\'e 
averaged 6th order WKB approximation method to obtain the QNMs numerically.

Taking into consideration of axial perturbation only, it is possible to express the perturbed metric in the following form \cite{Bouhmadi-Lopez:2020oia, Gogoi:2023kjt}:
\begin{align} \label{pert_metric}
ds^2 =& -\, |g_{tt}|\, dt^2 + r^2 \sin^2\!\theta\, (d\phi - p_1(t,r,\theta)\,
dt \notag \\ &- p_2(t,r,\theta)\, dr - p_3(t,r,\theta)\, d\theta)^2 + g_{rr}\, dr^2 \notag \\ & +
r^2 d\theta^2.
\end{align}
In this context, the parameters $p_1$, $p_2$, and $p_3$ are intricately linked to the perturbation occurring in the spacetime surrounding the black hole. These parameters play a crucial role in characterizing the modifications induced by the perturbation. Specifically, they contribute to the altered dynamics and structure of the black hole spacetime.

Now, for the massless scalar field, as it is considered that the 
effect of the field on the spacetime is minimal, the perturbed 
metric Eq. \eqref{pert_metric} takes the following form: 
\begin{equation}
ds^2 = -\,|g_{tt}|\, dt^2 + g_{rr}\, dr^2 +r^2 d \Omega^2.
\end{equation}

The Klein-Gordon equation in curved spacetime in this case is written as
\begin{equation}  \label{scalar_KG}
\square \Phi = \dfrac{1}{\sqrt{-g}} \partial_\mu (\sqrt{-g} g^{\mu\nu}
\partial_\nu \Phi) = 0.
\end{equation}
Eq. \eqref{scalar_KG} describes the QNMs associated with the scalar perturbation in the considered black hole spacetime. 

Utilising spherical harmonics, it is possible to express the wave function of the scalar field in the following decomposed form:

\begin{equation}\label{scalar_field}
\Phi(t,r,\theta, \phi) = \sum_{l,m} \dfrac{\psi_l(r)}{r} Y_{lm}(\theta,
\phi)e^{-i\omega t} .
\end{equation}

In the above equation, $Y_{lm}(\theta,\phi)$ is the spherical harmonics part of the wave function and the parameters $l$ and $m$ are the associated parameters of spherical harmonics. Now, one can use this decomposed form of the wave function in the Eq.  \eqref{scalar_field} to obtained a reduced wave equation given by

\begin{equation}  \label{radial_scalar}
\partial^2_{r_*} \psi_l(r_*) + \omega^2 \psi_l(r_*) = V_s(r) \psi_l(r_*).
\end{equation}
This is a standard differential equation where the differentiation is done with respect to the tortoise coordinate $r_*$. The tortoise coordinate is connected with the normal coordinate $r$ by the following definition: 
\begin{equation}  \label{tortoise}
\dfrac{dr_*}{dr} = \sqrt{g_{rr}\, |g_{tt}^{-1}|}.
\end{equation}

\begin{figure}[ht!]
      	\centering{
      	\includegraphics[scale=0.75]{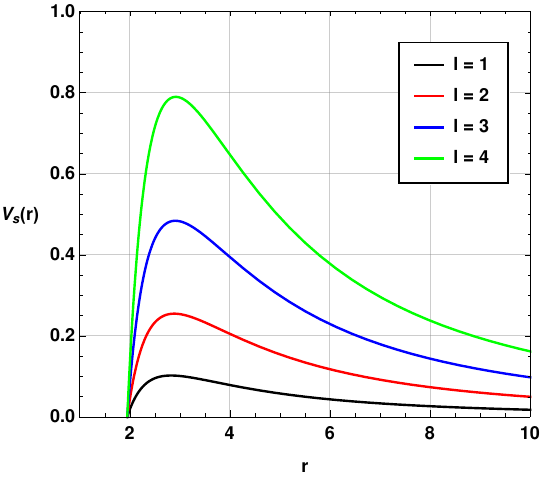}}
      	\caption{Variation scalar potential with respect to $r$ using $M = 1$, $Q = 0.3$ and $\xi = 0.1$.}
      	\label{figPot01}
      \end{figure}

In the Eq. \eqref{radial_scalar}, parameter $V_s(r)$ 
is the effective potential for the massless scalar perturbation and it is given by: 
\begin{equation}  \label{Vs}
V_s(r) = |g_{tt}| \left( \dfrac{l(l+1)}{r^2} +\dfrac{1}{r \sqrt{|g_{tt}|
g_{rr}}} \dfrac{d}{dr}\sqrt{|g_{tt}| g_{rr}^{-1}} \right).
\end{equation}
In this context, the symbol $l$ is employed to signify the multipole moment associated with the QNMs of the black hole.  The variation of the scalar potential $V_s(r)$ with respect to the multipole moment $l$ is shown in Fig. \ref{figPot01}. The overall effective potential increases with $\ell$.

%\begin{widetext}
      \begin{figure*}[t!]
      	\centering{
      	\includegraphics[scale=0.75]{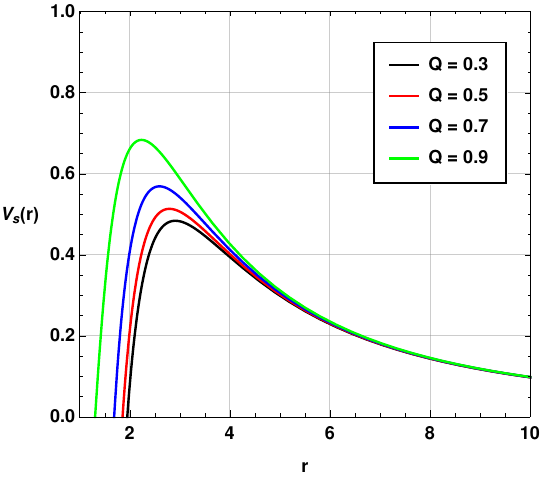} \hspace{0.2cm}
       \includegraphics[scale=0.75]{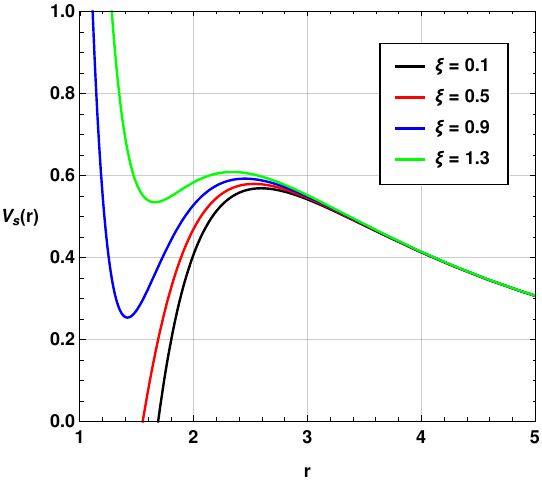}}
      	\caption{Variation scalar potential with respect to $r$ using $M = 1$ and $l = 3$. On the first panel, we have used $\xi = 0.1$ and on the second $Q = 0.7$. For $\xi=0.9$ and $\xi=1.3$, these configurations are not black holes but represent solitonic solutions. }
      	\label{figPot02}
      \end{figure*}  
%\end{widetext}

The other model parameters $Q$ and $\xi$ also have noticeable impacts on the potential $V_s(r)$ which are depicted in Fig. \ref{figPot02}. It is seen that with an increase in the black hole charge $Q$, the peak value of the potential increases non-linearly and the corresponding $r$ shifts towards the black hole event horizon. Similarly, an increase in the value of $\xi$ also increases the peak value of the potential. However, the variation is comparatively smaller at the peak position and the peak value seems to change uniformly with respect to $\xi$. This suggests that the model parameter $Q$ might have a non-linear impact on the spectrum of QNMs and $\xi$ might have an almost linear impact on QNMs. The details will be investigated in the next part of the study. 

\subsection{The Pad\'e averaged WKB approximation method}

In this study, we implement a well-known method called WKB approximation method to calculate the QNMs numerically for the massless scalar perturbation in the above-mentioned black hole spacetime. We consider Pad\'e averaged 6th-order corrections to the WKB method \cite{Schutz:1985km, Iyer:1986np, Konoplya:2003ii, Matyjasek:2019eeu} to obtain QNMs with higher accuracy. 

In higher-order WKB approximation method, the oscillation frequency $\omega$ of ring-down GWs is expressed as:
\begin{equation}
\omega = \sqrt{-\, i \left[ (n + 1/2) + \sum_{k=2}^6 \bar{\Lambda}_k \right] \sqrt{-2 V_0''} + V_0}.
\end{equation}
Here, the parameter $n$, representing the overtone number, takes on discrete values starting from zero $(n = 0, 1, 2, \hdots)$. Additionally, $V_0$ represents the value of the potential function $V(r)$ at a specific radial coordinate $r_{\text{max}}$, while $V_0''$ signifies the second derivative of $V(r)$ with respect to $r$ evaluated at the same position. The pivotal location $r_{\text{max}}$ corresponds to where the potential function achieves its maximum value, rendering this radial coordinate particularly significant.

Integral to our computational framework is the correction terms $\bar{\Lambda}_k$, which play a crucial role in refining the precision of our calculations. These correction terms are instrumental in capturing subtle nuances in the determination of the oscillation frequency, contributing to the overall accuracy of the framework.

To elaborate on the procedural aspects of our approach, we draw upon established references in the field. Specifically, Ref.s \cite{Schutz:1985km, Iyer:1986np, Konoplya:2003ii, Matyjasek:2019eeu} provide comprehensive insights into the mathematical formulations of the correction terms and elaborate on the intricacies of the Pad\'e averaging method applied in our study.

By incorporating these computational techniques, our research aims to uncover the nuanced behaviour of ring-down GWs, making significant contributions to the broader understanding of astrophysical phenomena in such types of theories of gravity.

\begin{widetext}

\begin{table}[htb!]
{\centering
\begin{center}
\caption{QNMs with different values of multipole moment $l$ calculated using $6$th order Pad\'e averaged WKB approximation method. In
this table we have chosen $n = 0, M=1, Q = 0.3$ and $\xi = 0.1$. }
\vspace{0.3cm}
\begin{tabular}{cccc}
\hline\\[-10pt]
%\multicolumn{1}{|l}{$l$} & \multicolumn{1}{l}{$M(\omega_R - i \omega_I)$} & \multicolumn{1}{l}{$\Delta_{rms}$} & \multicolumn{1}{l|}{$\Delta_6$}  \\
$l$ & Pad\'e averaged WKB & $\Delta_{rms}$ & $\Delta_6$ \\[2pt]  \hline
\\[-10pt] 

 $1$ & $0.297616 - 0.098004 i$ & $1.18328 \times 10^{-9}$ & $0.0000209617$ \\
 $2$ & $0.491302 - 0.097122 i$ & $3.04799\times 10^{-7}$ & $2.38347\times 10^{-6}$ \\
 $3$ & $0.686034 - 0.096870 i$ & $4.63924\times 10^{-8}$ & $4.55756\times 10^{-7}$ \\
 $4$ & $0.881104 - 0.096766 i$ & $1.04331\times 10^{-8}$ & $1.20843\times 10^{-7}$ \\
 $5$ & $1.076325 - 0.096713 i$ & $3.41801\times 10^{-9}$ & $4.44961\times 10^{-8}$ \\
 $6$ & $1.271628 - 0.096682 i$ & $1.36926\times 10^{-9}$ & $1.88337\times 10^{-8}$ \\
 $7$ & $1.466978 - 0.096663 i$ & $6.75515\times 10^{-10}$ & $9.20866\times 10^{-9}$ \\
 \hline  
\end{tabular}
\label{Table01}
\end{center}}
\end{table}
\end{widetext}
Table \ref{Table01} showcases the QNMs across a range of multipole moment values, with a specific focus on instances where the overtone number is set to $n=0$. The second column of the table delineates the QNMs obtained through the application of the 6th-order Pad\'e averaged WKB approximation method. Within this tabular representation, the term $\Delta_{rms}$ assumes significance as it quantifies the root mean square error associated with the 6th-order Pad\'e averaged WKB approximation.

Moreover, the term $\Delta_6$ is introduced to gauge the error between two adjacent order approximations. This error term is computed by assessing the absolute difference between $\omega_7$, representing QNMs computed using the Pad\'e averaged 7th order WKB approximation method, and $\omega_5$, denoting QNMs obtained from the Pad\'e averaged 5th order WKB approximation method. Mathematically, $\Delta_6$ is expressed as follows:
\begin{equation}
\Delta_6 = \dfrac{|\omega_7 - \omega_5|}{2}.
\end{equation}
 A notable trend within the table is the observed reduction in error associated with QNMs as the multipole moment $l$ increases. This behaviour aligns with the characteristic limitations of the WKB approximation method, particularly evident when the overtone number $n$ surpasses the multipole moment $l$. The diminishing error with increasing multipole moment is a distinctive feature of the WKB method, signifying its challenges in providing accurate results when overtone numbers exceed the corresponding multipole moment values \cite{Lambiase:2023hng, sekhmani_electromagnetic_2023, Gogoi:2023kjt, Parbin:2022iwt, Gogoi:2022ove, karmakar_quasinormal_2022, Gogoi:2022wyv}.
By keeping this fact in mind, in this investigation, we have considered $n<l$ to calculate QNMs.

\begin{figure*}[t!]
      	\centering{
      	\includegraphics[scale=0.55]{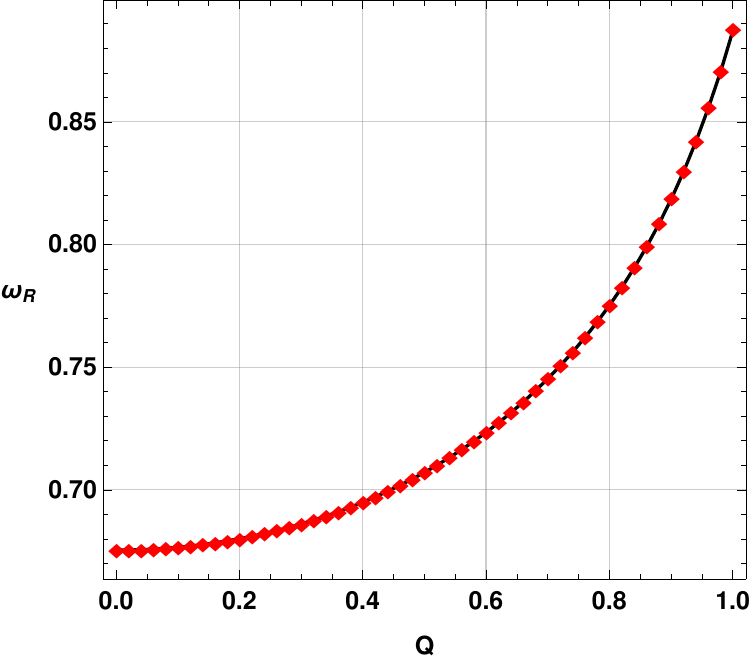}
       \includegraphics[scale=0.55]{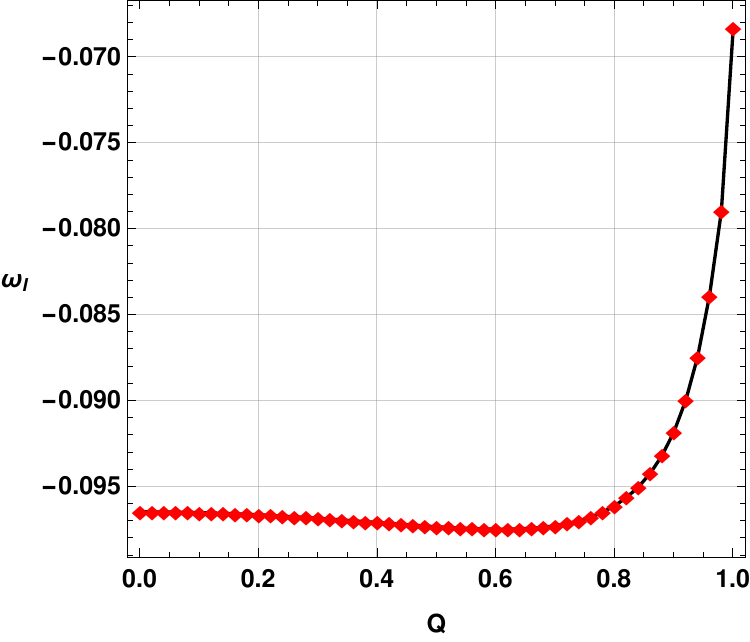}}
      	\caption{Variation of real and imaginary QNMs using $M = 1$, $l = 3$ and $\xi = 0.1$. }
      	\label{figQNM01}
      \end{figure*}

      \begin{figure*}[t!]
      	\centering{
      	\includegraphics[scale=0.55]{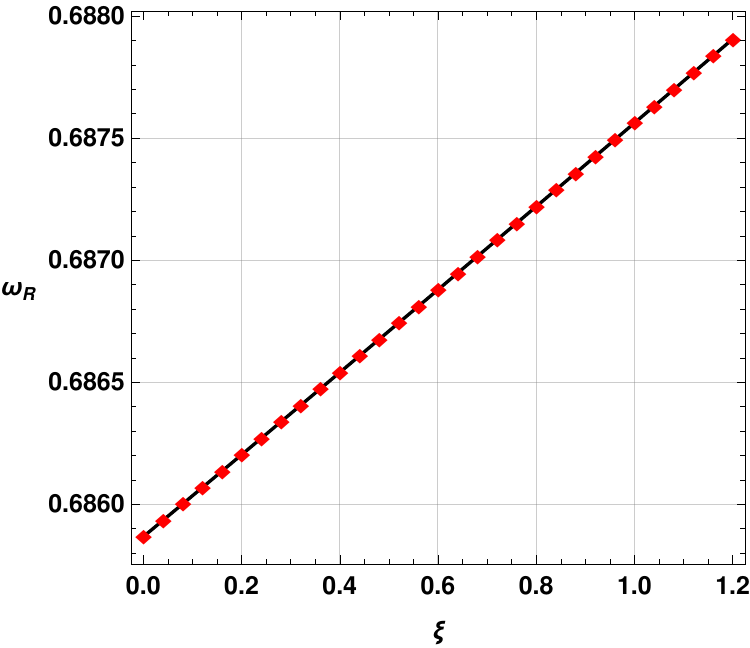}
       \includegraphics[scale=0.56]{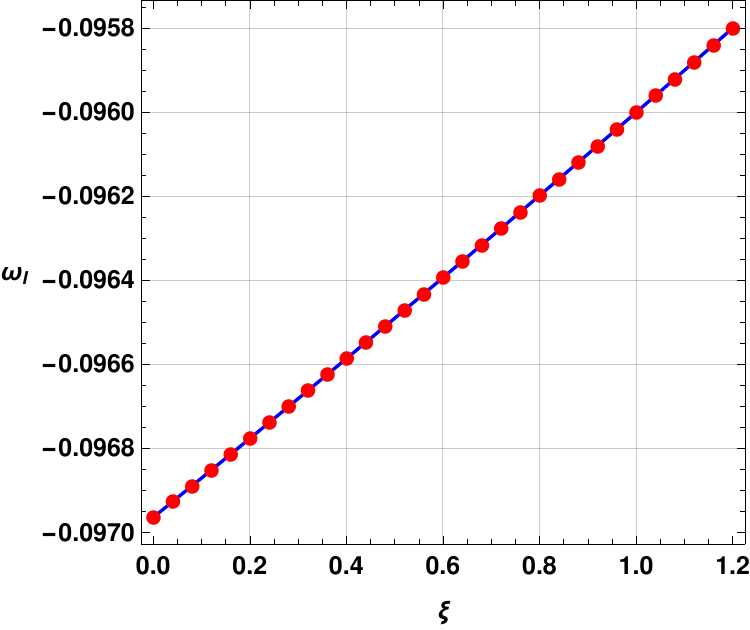}}
      	\caption{Variation of real and imaginary QNMs using $M = 1$, $l = 3$ and $Q = 0.3$.}
      	\label{figQNM02}
      \end{figure*}

We have shown the variation of the massless scalar QNMs in Fig. \ref{figQNM01} and \ref{figQNM02} with respect to $Q$ and $\xi$, respectively. In Fig. \ref{figQNM01}, one can see that with an increase in the charge of the black hole $Q$, the real QNMs increase non-linearly. On the other hand, the damping rate or decay rate of ring-down GWs increases very slowly initially up to around $Q=0.6$ and beyond this the damping rate decreases drastically showing a highly non-linear pattern.

In Fig. \ref{figQNM02}, the initial panel intricately illustrates the nuanced shift in real quasinormal frequencies concerning the parameter $\xi$, while the second panel meticulously delineates the corresponding alteration in the damping rate of ring-down GWs with respect to $\xi$. Evidently, as the parameter $\xi$ experiences an incremental surge, the real quasinormal frequencies manifest a discernible linear escalation. In contrast, a rise in the value of $\xi$ induces a linear reduction in the damping rate. It is noteworthy, however, that this variation is comparatively modest when juxtaposed with the scenario involving the parameter $Q$.

Consequently, when contemplated through the prism of QNMs, the influences exerted by $Q$ and $\xi$ exhibit notable disparities. Looking forward, as we anticipate a wealth of data from state-of-the-art GW detectors such as Laser Interferometer Space Antenna (LISA), there emerges the exciting prospect of establishing stringent constraints on the Yang-Mills field by harnessing insights derived from the meticulous examination of QNMs.

\section{Time Evolution of Scalar Perturbations} \label{sec04}

In the previous section, we conducted numerical computations to unravel the characteristics of QNMs and scrutinized how they hinge on the model parameters. Shifting our attention to the subsequent section, our exploration now centers on the temporal behaviours of massless scalar perturbations. To elucidate these profiles, we adopt the time domain integration framework delineated by Gundlach et al. \cite{Gundlach:1993tp,Gundlach:1993tn}

To progress further, we denote $\psi(r^*,
t) = \psi(i \Delta r^*, j \Delta t) = \psi_{i,j} $, and $V(r(r^*)) =
V_{i}$. Here $r^*$ represents the tortoise coordinate and $t$ denotes time. Subsequently, we express the scalar wave equation \eqref{scalar_KG} as:

\begin{align}
\dfrac{\psi_{i+1,j} - 2\psi_{i,j} + \psi_{i-1,j}}{\Delta r_*^2} - &\dfrac{\psi_{i,j+1} - 2\psi_{i,j} + \psi_{i,j-1}}{\Delta t^2} \notag \\  &- V_i\psi_{i,j} = 0.
\end{align}

The initial conditions are specified as $\psi(r_*,t) = \exp \left[ -\dfrac{(r_* - k_1)^2}{2\sigma^2} \right]$ and $\psi(r_*,t)\vert_{t<0} = 0$, where $k_1$ and $\sigma$  represent the median and width of the initial wave-packet, respectively. Employing these initial conditions, we compute the time evolution associated with the scalar perturbation using the iterative scheme:
\begin{widetext}
 \begin{equation}
\psi_{i,j+1} = -\,\psi_{i, j-1} + \left( \dfrac{\Delta t}{\Delta r_*}
\right)^2 (\psi_{i+1, j + \psi_{i-1, j}}) + \left( 2-2\left( \dfrac{\Delta t%
}{\Delta r_*} \right)^2 - V_i \Delta t^2 \right) \psi_{i,j}.
\end{equation}   
\end{widetext}

We implement this iterative scheme to obtain the time domain profiles by keeping a fixed value of
 $\frac{\Delta t}{\Delta r_*}$. However, it is important to mention that 
one must consider $\frac{\Delta t}
{\Delta r_*} < 1$ to satisfy the Von Neumann stability 
condition throughout the numerical procedure. Throughout our numerical calculations, we have considered this ratio around $0.7$ to obtain the time evolution of the massless scalar perturbations on the black hole spacetime. We have set $k_1=0$ and $\sigma=5$ for the numerical calculations of time domain profiles.

      \begin{figure*}[t!]
      	\centering{
      	\includegraphics[scale=0.75]{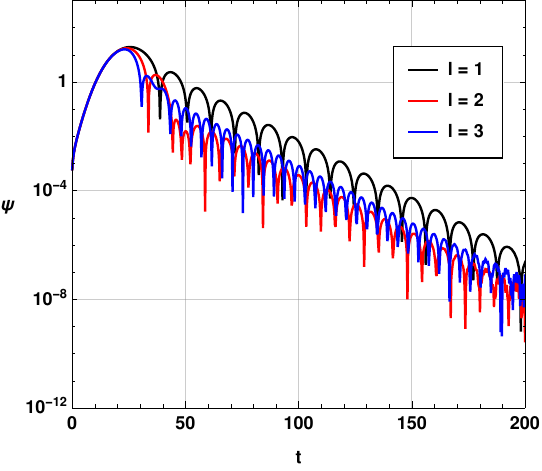}
       \includegraphics[scale=0.75]{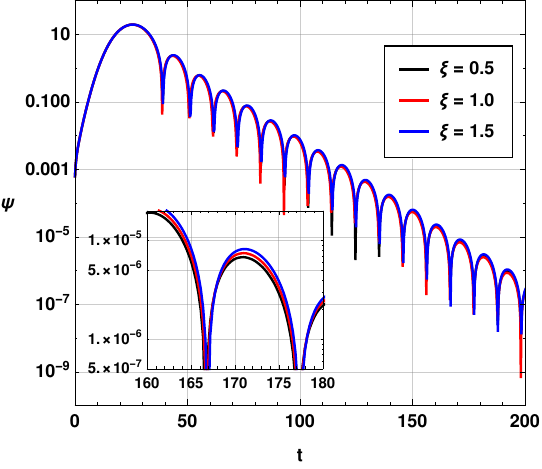}
       \includegraphics[scale=0.75]{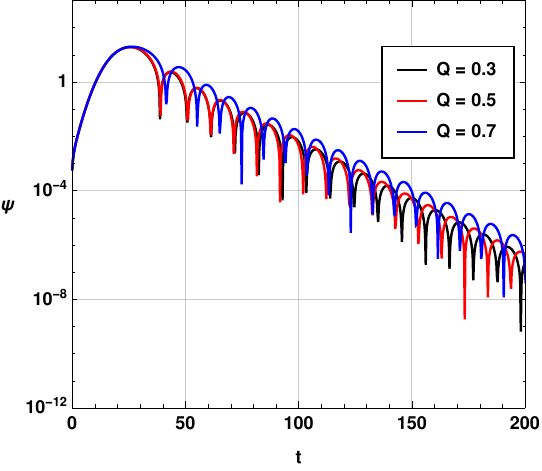}}
      	\caption{Time domain profiles of scalar perturbation with $M = 1$. On the first panel, $Q = 0.3$ and $\xi = 0.5$; on the second panel $l = 1$ and $Q = 0.3$ and on the third panel $l=1$ and $\xi = 0.5$.}
      	\label{figtd01}
      \end{figure*}

In Fig. \ref{figtd01}, we have shown the time domain profiles in the units of $M$. One may observe that the variation of time domain profiles with respect to multiple moment $l$ stands in agreement with our Table \ref{Table01}. From the second and third panel, it is evident that the impact of the coupling parameter $\xi$ on the QNMs spectrum is very small in comparison to the charge parameter $Q$ as predicted earlier. These results are also in agreement with the results obtained from Pad\'e averaged WKB approximation method.

\section{Optical Behaviour of the Black hole: Shadow} \label{sec05}

The investigation into supermassive black holes (SMBHs) holds significant importance in astrophysics, given their mysterious nature and their central positions in galaxies, where they exert exceptional gravitational forces with event horizons preventing light escape. Recent advances in observational techniques, exemplified by the Event Horizon Telescope (EHT), have offered unprecedented insights into these colossal cosmic entities.

In this part, we explore the mysterious concept of black hole shadows, uncovering the hidden areas within the complex structure of space and time that cover the edge of these cosmic objects. Black holes have incredibly strong gravity that forces even light to be pulled in when it crosses the boundary called the event horizon \cite{EslamPanah:2020hoj, Konoplya2019}. Consequently, the outer boundary of a black hole manifests as an indistinct yet solemn circle, projecting its inscrutable silhouette against the surrounding cosmic tableau of matter or luminous radiance. The dimensions and intricate contours of this umbral figure furnish profound insights into the fundamental characteristics of black holes and the essence of gravity itself. Recent advancements in both astronomical observation and technological capabilities have conferred upon us the capacity to capture and visually scrutinize these elusive black hole shadows, representing a paradigmatic leap in our pursuit to comprehend these enigmatic celestial phenomena \cite{gogoi_joulethomson_2023}. As we peer into these cosmic shadows, we not only unveil the visual facets of black holes but also unlock new strata of scientific comprehension, shedding light on the intricacies veiled within the expansive cosmic tapestry. These advancements herald a gateway to a deeper understanding of the universe's intricacies, underscoring the remarkable synergy between cutting-edge technology and our relentless scientific inquisitiveness about the cosmos. Such a study will extensively contribute towards our understanding of different modified gravity theories and the validity of GR in light of current observational aspects.

The Euler-Lagrange equation is given by
\begin{equation}\label{shadow0}
\frac{d}{d\tau}\!\left(\frac{\partial\mathcal{L}}{\partial\dot{x}^{\mu}}\right)-\frac{\partial\mathcal{L}}{\partial x^{\mu}}=0,
\end{equation}
where the Lagrangian is expressed as
\begin{equation}\label{shadow1}
\mathcal{L}(x,\dot{x})=\frac{1}{2}\,g_{\mu\nu}\dot{x}^{\mu}\dot{x}^{\nu}.
\end{equation}
For static and spherically symmetric black hole,
the Lagrangian becomes
\begin{equation}\label{shadow2}
\mathcal{L}(x,\dot{x})=\frac{1}{2}\left[-f(r)\,\dot{t}^{2}+\frac{1}{f(r)}\,\dot{r}^{2}+r^{2}\left(\dot{\theta}^{2}+\sin^{2}\theta\dot{\phi}^{2}\right)\right].
\end{equation}
Here the dot over the variables denotes the derivative with respect to the proper time $\tau$.

Choosing the equatorial plane where $\theta=\pi/2$, the conserved energy $\mathcal{E}$ and angular momentum $L$ can be computed by utilizing the killing vectors $\partial/\partial \tau$ and $\partial/\partial \phi$ as given by
\begin{equation}
\mathcal{E}=f(r)\,\dot{t},\quad L=r^{2}\dot{\phi}.
\end{equation}
In the case of photon, we can write the geodesic equation as,
\begin{equation}\label{eq22}
-f(r)\,\dot{t}^{2}+\frac{\dot{r}^{2}}{f(r)}\,+r^{2}\dot{\phi}^{2} = 0.
\end{equation}
Utilizing Eq. \eqref{eq22} in conjunction with the conserved quantities, namely, $\mathcal{E}$ and $L$, the orbital equation for a photon can be derived as. \cite{gogoi_joulethomson_2023}
\begin{equation}\label{eff}
\left(\frac{dr}{d\phi}\right)^{2}=V_{eff},
\end{equation}
where $V_{eff}$ is defined as
\begin{equation}
V_{eff}= r^{4} \left[\frac{\mathcal{E}^{2}}{L^{2}}-\frac{f(r)}{r^{2}}\right].
\end{equation}
Expressing Eq. \eqref{eff} in radial form yields,
\begin{equation}
    V_r(r) = \dfrac{1}{\zeta^2} - \dot{r}^2/L^2.
\end{equation}
In this expression, $\zeta$ represents the impact parameter and is defined as $\zeta=L/\mathcal{E}$. Additionally, the reduced potential $V_r(r)$ can be expressed as:
\begin{equation}\label{pot}
V_r(r) = \frac{f(r)}{ r^2}.
\end{equation}

To scrutinize the shadows of black holes, our focus is directed towards a specific point along the trajectory denoted as $r_{ph}$. This particular point corresponds to the turning point, signifying the location of the light ring that encircles the black hole. It can also be interpreted as the radius of the photon sphere, holding significant relevance in the analysis of the black hole's shadow characteristics. At this pivotal turning point of a black hole
\cite{Jafarzade:2020ova, EslamPanah:2020hoj, pantig_shadow_2022, Ovgun:2018tua, Papnoi:2014aaa, Ovgun:2019jdo, gogoi_joulethomson_2023},
\begin{equation}
\left.V_{eff}\right|_{r_{ph}}\!\!\!\!=0,\;\; \text{and}\;\; 
\left.V_{eff}\right|_{r_{ph}}\!\!=0.
\end{equation} 
It is feasible to determine the impact parameter $\zeta$ at the turning point by
\begin{equation}
\frac{1}{\zeta_{crit}^{2}}=\frac{f(r_{ph})}{r_{ph}^{2}}.
\label{impact}
\end{equation}
Therefore, the radius of the photon sphere, denoted as $r_{ph}$, can be determined using\cite{gogoi_joulethomson_2023}: 
\begin{equation}
\left.\frac{d}{dr}\,\mathcal{A}(r)\right|_{r_{ph}}\!\!\!\!\!\!\! = 0.
\end{equation}
This equation can be reformulated as
\begin{equation}
\frac{f^{\prime}(r_{ph})}{f(r_{ph})}-\frac{h^{\prime}(r_{ph})}{h(r_{ph})}=0,
\label{photon}
\end{equation}
here we have $\mathcal{A}(r)=h(r)/f(r)$ with $h(r)=r^{2}$. 

Now, to derive the shadow of the black hole, we express Eq. \eqref{eff} using Eq. \eqref{impact} in terms of the function $\mathcal{A}(r)$ as 
\begin{equation}
\left(\frac{dr}{d\phi}\right)^{\!2}= h(r)f(r)\left(\frac{\mathcal{A}(r)}{\mathcal{A}(r_{ph})}-1\right).
\label{eq33}
\end{equation}
Utilizing Eq. \eqref{eq33}, one can calculate the shadow radius of the black hole. In the scenario where a static observer is positioned at a distance $r_0$ from the black hole, we can ascertain the angle $\alpha$ between the light rays originating from the observer and the radial direction of the photon sphere. This angle can be computed as \cite{gogoi_joulethomson_2023}
\begin{equation}
\cot\alpha=\frac{1}{\sqrt{f(r)h(r)}}\left.\frac{dr}{d\phi}\right|_{r\,=\,r_{0}}\!\!\!\!\!\!\!\!\!\!\!.
\end{equation}

In conjunction with Eq. \eqref{eq33}, the aforementioned equation can be represented as 
\begin{equation}
\cot^{2}\!\alpha=\frac{\mathcal{A}(r_{0})}{\mathcal{A}(r_{ph})}-1.
\end{equation}
Once more, the above equation can be reformulated using the relation $\sin^{2}\!\alpha=1/(1+\cot^{2}\!\alpha)$ as
\begin{equation}
\sin^{2}\!\alpha=\frac{\mathcal{A}(r_{ph})}{\mathcal{A}(r_{0})}.
\end{equation}

By substituting the expression for $\mathcal{A}(r_{ph})$ from Eq. \eqref{impact} and using $\mathcal{A}(r_{0}) = r_0^2/f(r_0)$, the shadow radius of the black hole for a static observer at $r_{0}$ can be approximated as \cite{gogoi_joulethomson_2023} 
\begin{equation}
R_{s}=r_{0}\sin\alpha=\sqrt{\frac{r_{ph}^2f(r_{0})}{f\left(r_{ph}\right)}}.
\end{equation}
Once more, as $r_0 \rightarrow \infty$, i.e., for a static observer at a large distance, $f(r_0) \rightarrow 1$. Consequently, for such an observer, the shadow radius $R_s$ becomes,
\begin{equation}
R_{s} = \frac{r_{ph}}{\sqrt{f(r_{ph})}}.
\end{equation}
Finally, through the stereographic projection of the shadow from the black hole's plane to the observer's image plane with coordinates $(X,Y)$, the apparent form of the shadow can be determined. These coordinates are defined as \cite{gogoi_joulethomson_2023, karmakar_thermodynamics_2023, Anacleto:2021qoe} 
\begin{align}
 X & =\lim_{r_{0}\rightarrow\infty}\left(-\,r_{0}^{2}\sin\theta_{0}\left.\frac{d\phi}{dr}\right|_{r_{0}}\right),\\[5pt]
Y & =\lim_{r_{0}\rightarrow\infty}\left(r_{0}^{2}\left.\frac{d\theta}{dr}\right|_{(r_{0},\theta_{0})}\right).
\end{align}
In this context, $\theta_{0}$ denotes the angular orientation of the observer in relation to the plane of the black hole.

\begin{figure*}[t!]
      	\centering{
      	\includegraphics[scale=0.65]{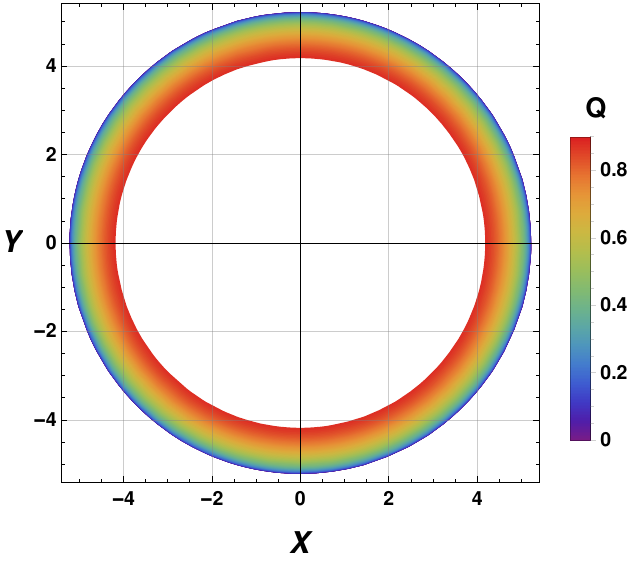}
       \includegraphics[scale=0.65]{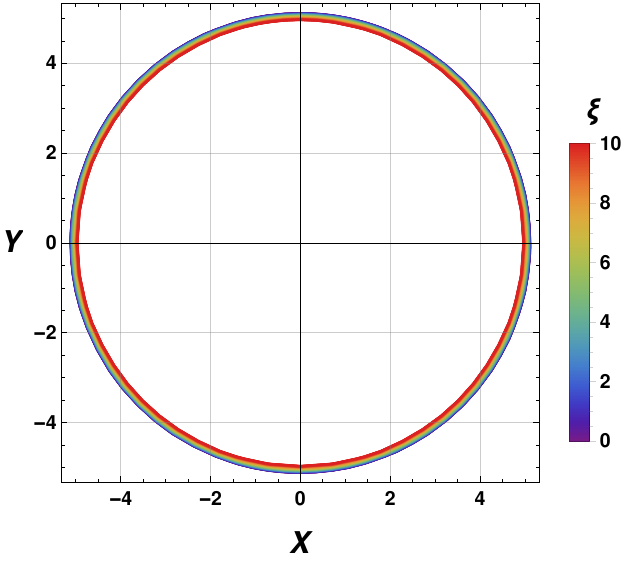}}
      	\caption{Stereographic projection of black hole shadow using $M = 1$. On the first panel, we have used $Q = 0.3$ and on the second panel $\xi = 0.5$. }
      	\label{figSh01}
      \end{figure*}

\begin{figure*}[t!]
      	\centering{
      	\includegraphics[scale=0.75]{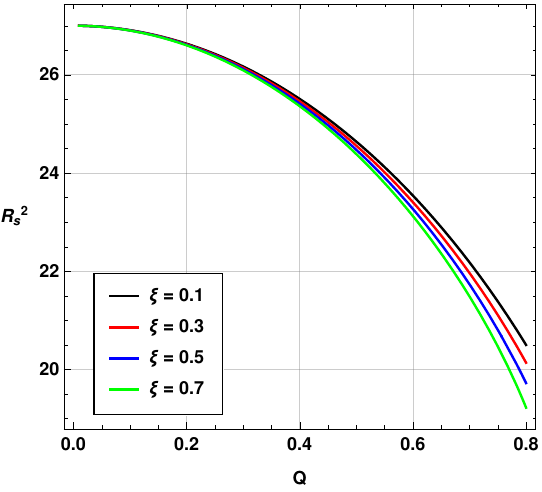}
       \includegraphics[scale=0.37]{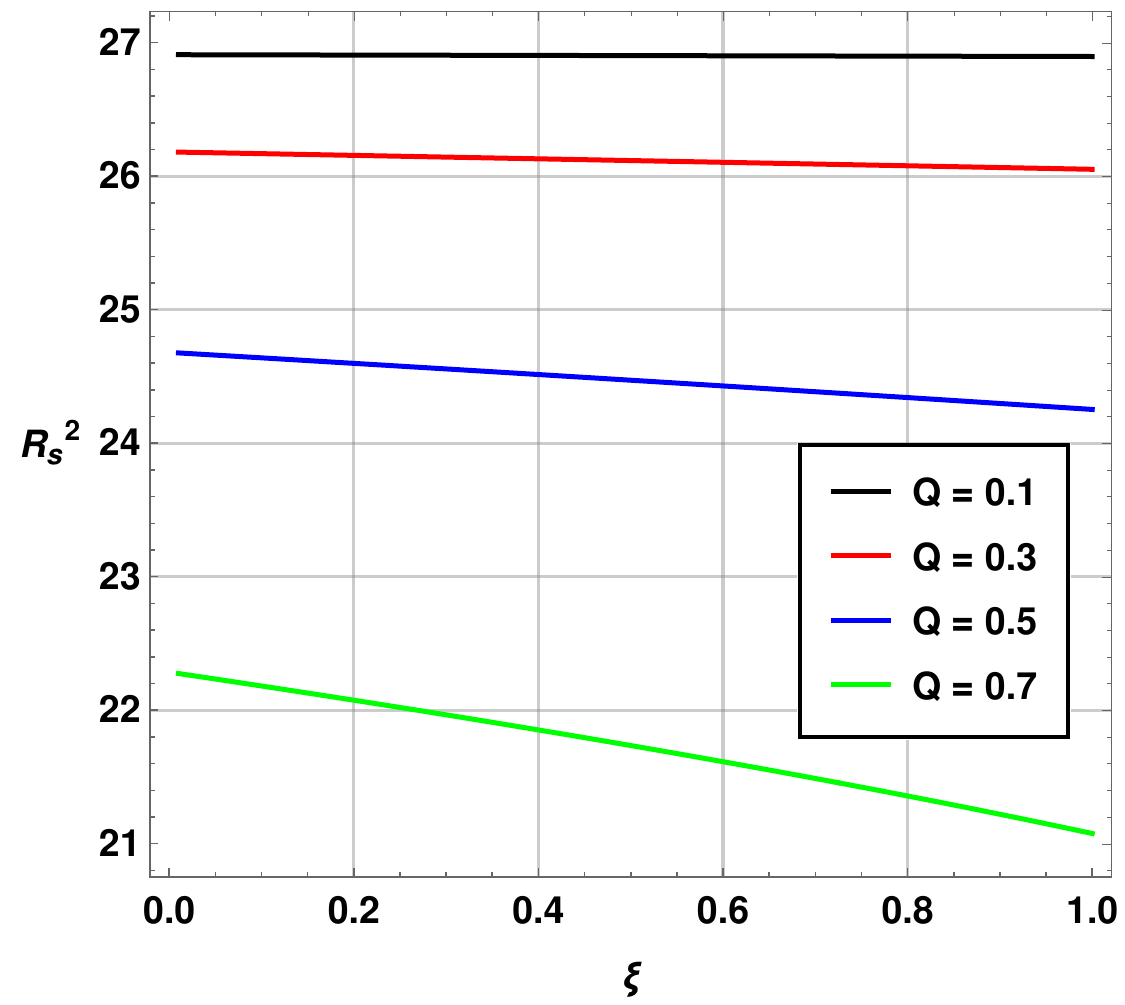}}
      	\caption{Variation of $R_s^2$ as a function of black hole charge $Q$ (the left panel) and non-minimal coupling constant $\xi$ (the right panel). We have used $M = 1$. } 
      	\label{figSh02}
      \end{figure*}

      \begin{figure*}[t!]
      	\centering{
      	\includegraphics[scale=0.65]{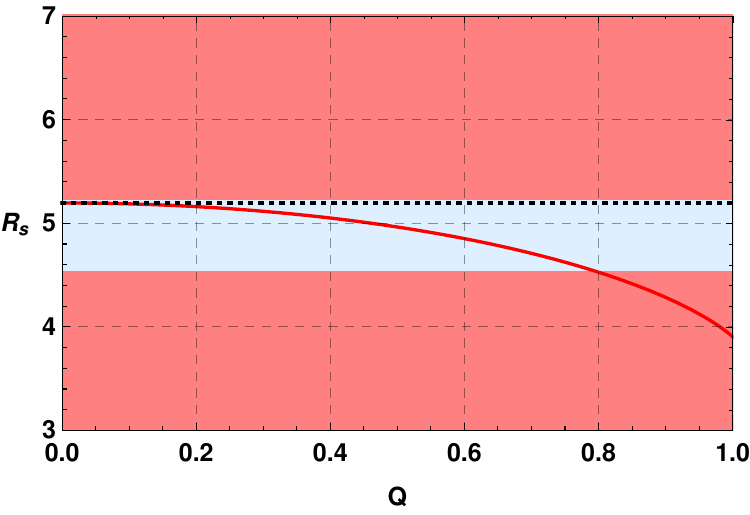}
       \includegraphics[scale=0.65]{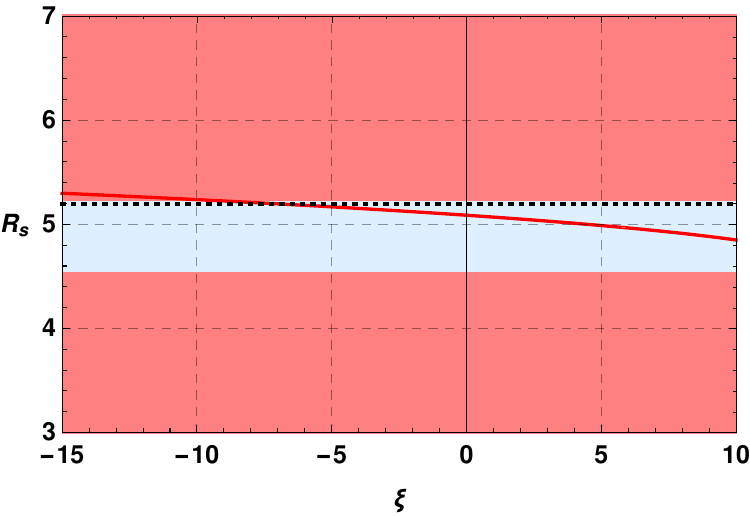}}
      	\caption{Constraint on black hole shadow radius from Sgr A*. On the left panel, we have used $\xi = 0.1$ and on the right panel, we have used $Q = 0.35$. The excluded (allowed) regions are denoted by the red (grey) shaded areas. The black dotted horizontal line is the radius of the shadow of the Schwarzschild black hole.} 
      	\label{figSh03}
      \end{figure*}

      \begin{figure*}[t!]
      	\centering{
      	\includegraphics[scale=0.65]{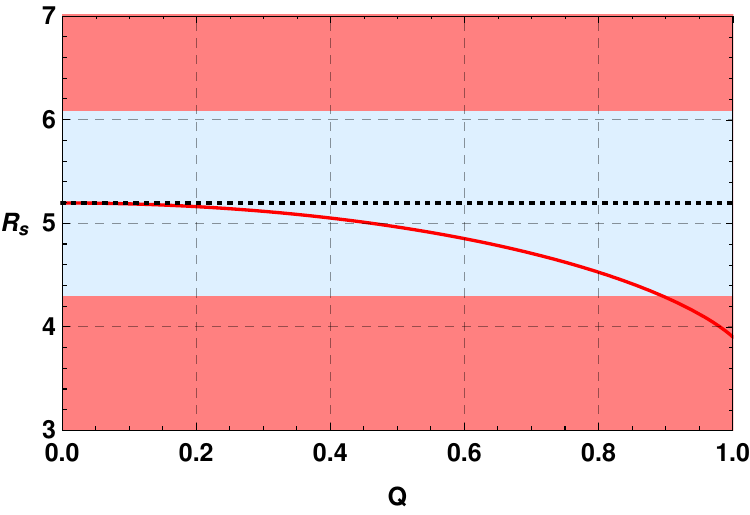}
       \includegraphics[scale=0.65]{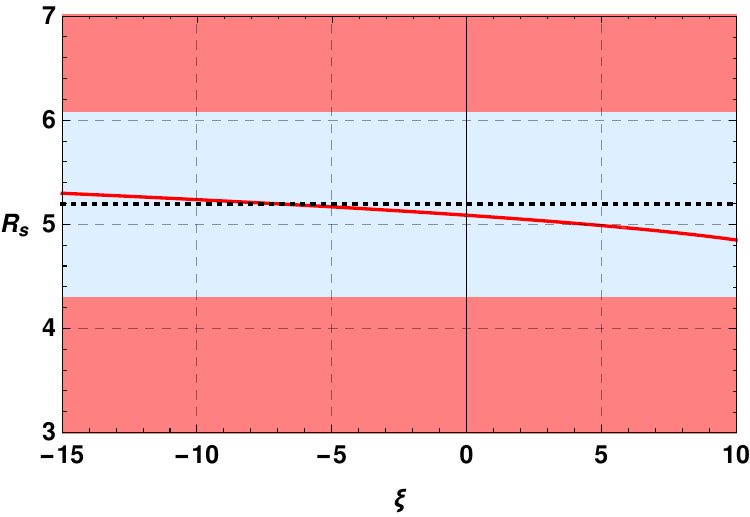}}
      	\caption{Constraint on black hole shadow radius from M87*. On the left panel, we have used $\xi = 0.1$ and on the right panel, we have used $Q = 0.35$. The excluded (allowed) regions are denoted by the red (grey) shaded areas. The black dotted horizontal line is the radius of the shadow of the Schwarzschild black hole.} 
      	\label{figSh04}
      \end{figure*}

\begin{table*}[!ht]
    \centering
    \begin{tabular}{ccc}
\hline
\hline
 &   constraint on $Q$ using $\xi =0.1$ \;\; &  constraint on $\xi$ using $Q = 0.35$  \\
\hline
$1\sigma$(upper/lower) from M87*    &       none /$0.8912$ &       $-138.11$/none  \\
$1\sigma$(upper/lower) from Sgr A*    & none/$0.789$ &$ -8.695$/none \\
\hline
\end{tabular}
\caption{Variation of shadow with $1\sigma$ ranges of $Q$ and $\xi$ based on the shadow radii of Sgr. A* and M87* as depicted in Figs. \ref{figSh03} and \ref{figSh04}.}
    \label{constrainttab}
\end{table*}

The investigation presents stereographic projections of the black hole shadow for the specific black hole under examination. Fig. \ref{figSh01} illustrates these projections, offering visual depictions of the shadow's characteristics corresponding to various model parameters. In the initial panel of Fig. \ref{figSh01}, the stereographic projections of the black hole are displayed for different values of the model parameters $Q$ and $\xi$. It is evident that an increase in the model parameter $Q$ results in a decrease in the black hole shadow. The second panel illustrates the black hole shadow for various values of the parameter $\xi$, revealing a gradual decrease in the shadow with an increase in $\xi$. For enhanced visualization, Fig. \ref{figSh02} presents the square of the black hole shadow radius concerning the model parameters $Q$ and $\xi$. This representation indicates that $\xi$ has a minimal impact on the black hole shadow. Therefore, the investigation demonstrates that both model parameters exert distinct influences on the black hole shadow.

A notable feature of SMBHs is the presence of a shadow cast by their event horizons, a phenomenon arising from the gravitational bending of light. The size and shape of this shadow furnish valuable information about the SMBH and its surrounding environment. To comprehend and quantify the shadow's dimensions, the spatial separation \(D\) between the SMBH and the galactic center must be considered.

A conventional method for determining the classical diameter of the black hole shadow involves the application of the arclength equation:

\begin{equation} \label{eqarc}
d_\text{sh} = \frac{D \theta_\text{sh}}{M},
\end{equation}

where \(d_\text{sh}\) denotes the shadow's diameter, \(\theta_\text{sh}\) represents the angular size of the shadow, and \(M\) corresponds to a specific unit of measurement.

For M87*, one of the extensively studied SMBHs, the measurement of the shadow diameter has been realized through EHT observations \cite{EventHorizonTelescope:2019dse, EventHorizonTelescope:2022wkp}. The reported diameter of the M87* shadow is \(d^\text{M87*}_\text{sh} = (11 \pm 1.5)M\), serving as a crucial data point for scrutinizing the properties of this SMBH.

However, it is imperative to consider the uncertainties associated with these measurements for a more accurate comprehension of the physical parameters. The determination of such uncertainties requires meticulous analysis, and in the case of the M87* shadow diameter, Ref.s \cite{EventHorizonTelescope:2021dqv, Vagnozzi:2022moj} offer detailed insights into the employed methodology.

This study aims to delve deeper into the implications of the measured M87* and Sgr A* shadow diameters by incorporating uncertainties from the aforementioned references. Our specific focus is on constraining the potential values of the black hole charge \(Q\) by calculating the \(1\sigma\) limits based on results obtained for M87* and Sgr A*. By considering these constraints, we aspire to augment our understanding of the model behaviour and the model parameters from the physical properties of M87*  and Sgr A* and contribute to the broader comprehension of SMBHs as well as possible constraints on QNMs from them.

In Figs. \ref{figSh03} and \ref{figSh04}, we have shown the constraints on the model parameters $Q$ and $\xi$ from the observed shadow radii of Sgr A* and M87* by following Ref.s \cite{EventHorizonTelescope:2021dqv, Vagnozzi:2022moj}. One can see that the observational data do not put a strong constraint on the coupling parameter $\xi$. However, the charge parameter $Q$ is well constrained by the observed shadow radii of Sgr A* and M87*.

\section{Connection between quasinormal modes and shadow of the black hole}
\label{sec06}

In this section, we shall briefly discuss the approximate analytical connection of QNMs with the black hole shadow radius.
One may note that from the numerical results of QNMs and black hole shadow, there is a possible correspondence between them.

In this scenario, we shall consider only $3^{rd}$ order WKB expansion given by 
\begin{widetext}
\begin{eqnarray}
\nonumber
\omega = \Bigg\{  V + \frac{V_4}{8V_2}\left( \nu^2+\frac{1}{4}\right) - \left( \frac{7 + 60\nu^2}{288}\right) \frac{V_3^2}{V_2^2}+i\nu \sqrt{-2V_2}\left[\frac{1}{2V_2}\left[\frac{5V_3^4(77+188\nu^2)}{6912V_2^4} \right. \right. \\
\nonumber
\left. \left. -\frac{V_3^2V_4(51+100\nu^2 )}{384V_2^3}+ \frac{V_4^2(67+68\nu^2 )}{2304V_2^2}+\frac{V_5V_3(19+28\nu^2 )}{288V_2^2}+\frac{V_6(5+4\nu^2 )}{288V_2}\right] -1 \right]  \Bigg\}^{1/2}_{r=r_{ph}}.\\
\label{e7}
\end{eqnarray}
\end{widetext}
One may note that even we consider higher order expansion such as $6$th order or $12$th order expansion, in the eikonal limit, the first significant terms are identical with those obtained by the $3$rd order WKB approximation method.
The term $V_i$ stands for the {\it i-th} derivative of the potential $V$. We also considered $\nu = n+\frac{1}{2}$, where $n$ is the overtone number of QNMs.  

By expanding \eqref{e7}, at the eikonal regime, one can obtain \cite{Gogoi:2023ffh, Cuadros-Melgar:2020kqn} 
\begin{widetext}
\begin{eqnarray}
\nonumber
\omega = \omega_R - i \omega_I = \left[\ell \frac{\sqrt{f(r)}}{r}\Bigg|_{r=r_{ph}} +  \frac{\sqrt{f(r)}}{2r}\Bigg|_{r=r_{ph}} + \mathcal{O}(\ell^{-1})\right]_R -\\
 i \left[\frac{n+1/2}{\sqrt{2}}\frac{\sqrt{f(r)}}{r}\Bigg|_{r=r_{ph}} 
\sqrt{6rf'(r)-6f(r)-r^2f''(r)-r^2f(r)^{-1}{f'(r)}^2}\Bigg|_{r=r_{ph}} +  \mathcal{O}(\ell^{-1})\right]_I.
\end{eqnarray} 
\end{widetext}
Comparing the imaginary part
\begin{equation}
\label{imgpart}
 \omega_I = \frac{n +1}{\sqrt{2}}R_s^{-1}\sqrt{2f(r_{ph})-r_{ph}^2f''(r_{ph})} +\mathcal{O}(\ell^{-1}),
\end{equation}
and the real part
\begin{equation}
\label{realpart}
\omega_R = R_s^{-1}\left(\ell + \frac{1}{2} +\mathcal{O}(\ell^{-1})\right).
\end{equation}

Further exploring the constraints on the visual representation of a black hole i.e. shadow can be achieved through the examination of observational data concerning its QNMs. It is essential to emphasize the direct relationship between the features of the black hole's silhouette and the specific values of its QNMs as this will help us to uncover different properties of the black hole hairs. Consequently, the availability of significant observational data from space-based gravitational wave detectors like LISA presents an avenue for imposing stricter constraints on the parameters of models used to understand black holes. Moreover, this process enables the assessment of consistency between two distinct approaches-shadow analysis and QNMs-in testing the underlying theoretical framework.

From the approximated relations \eqref{imgpart} and \eqref{realpart}, at the eikonal limit, we observe that the black hole shadow has an impact on the behaviour of the QNMs spectrum. Most importantly, an observational constraint on the black hole shadow will also put a constraint on the real and imaginary parts of QNMs of the black hole at the eikonal limit.
However, as evident from the table \ref{constrainttab}, such a constraint on the QNMs will be very weak at the current stage. In the near future, observational results from LISA might put a very strong constraint on QNMs which may be useful along with the observational results of black hole shadow to check for the consistencies between them and to constrain the theory more stringently.

\section{Concluding Remarks}\label{sec07}

In this work, we considered a charged black hole solution in non-minimally coupled EYM theory and studied the scalar perturbation on the black hole spacetime and the associated QNMs. From the behaviour of the potential, we found that for higher values of the coupling parameter $\xi$, the line element represents solitonic solutions instead of a black hole solution.

In order to obtain the QNMs, we used Pad\'e averaged WKB approximation method. The analysis showed that black hole charge $Q$ has a more significant impact on the quasinormal mode spectrum. With an increase in charge, the oscillation frequency of ring-down GWs increases non-linearly. In the case of the damping rate of GWs, we found that initially, with an increase in $Q$, the damping or decay rate increases slowly and beyond some threshold value around $Q=0.65$, it starts to decrease rapidly as $Q$ approaches $1$. One may note that some previous Ref.s \cite{Gogoi:2023ffh, Gogoi:2021cbp} also found similar variation trends of QNMs with respect to black hole charge $Q$. However, due to the presence of coupling term and non-linear charge distribution, in this case, we observe slightly different behaviour of QNMs towards higher values of $Q$.

Unlike black hole charge $Q$, the coupling parameter $\xi$ has a linear effect on the QNMs spectrum. With an increase in the value of $\xi$, the oscillation frequency of QNMs increases while the damping rate decreases slowly.

We also investigated time domain profiles for different model parameters and found that the profiles depict similar results as those found by using the WKB method.

In the next part of our work, we considered the shadow of the black hole and found that $Q$ has a noticeable impact on the shadow behaviour of the black hole. For a charged black hole, the shadow is expected to be smaller in size. A similar behaviour is observed in the case of $\xi$ also. However, the variation of shadow with respect to $\xi$ is small. Thus the impact of the coupling parameter as well as the presence of solitonic solutions may not be well differentiated by utilising shadow observations. As a result, observational data on black hole shadows, do not have a stringent constraint on $\xi$.

We also studied the relationship between the QNMs and the shadow of the black hole and found that the shadow observations can provide a constraint on the QNMs spectrum. In the near future, observational data of QNMs from LISA may provide us with scope to constrain the theory more efficiently and to check whether the shadow observations are well consistent with it or not.

Our study has uncovered that in the standard non-minimally coupled EYM theory, both the charge parameter $Q$ and the coupling term $\xi$ exert notable influences on the quasinormal spectrum and shadow of the black hole. The effects of these parameters on the quasinormal mode spectrum differ significantly in their nature.

\section*{Acknowledgments}
DJG acknowledges the contribution of the COST Action CA21136  -- ``Addressing observational tensions in cosmology with systematics and fundamental physics (CosmoVerse)". SP acknowledges the funding supported from the NSRF via the Program Management Unit for Human Resources $\&$ Institutional Development, Research and Innovation [grant number B37G 660013].

\section*{Data Availability Statement}
There are no new data associated with this article.

\bibliography{references}
\end{document}